\documentclass[pra,preprint,amsmath,amssymb]{revtex4}
\usepackage{epsfig,verbatim}

\newcommand{\curl}{\operatorname{curl}}
\newcommand{\Man}{\mathcal{M}}
\newcommand{\Ran}{\operatorname{Ran}}
\newcommand{\I}{\mathrm{i}}

\begin{document}

\title{Manifestly gauge invariant discretizations of the Schr\"odinger equation}
\date{\today}

\author{Tore Gunnar Halvorsen}
\email{t.g.halvorsen@cma.uio.no}
\affiliation{Centre of Mathematics for Applications, University of
  Oslo, N-0316 Oslo, Norway}

\author{Simen Kvaal}
\email{simen.kvaal@cma.uio.no}
\affiliation{Centre of Mathematics for Applications, University of
  Oslo, N-0316 Oslo, Norway}

\begin{abstract}
  Grid-based discretizations of the time dependent Schr\"odinger
  equation coupled to an external magnetic field are converted to
  manifest gauge invariant discretizations. This is done using
  generalizations of ideas used in classical lattice gauge theory, and
  the process defined is applicable to a large class of discretized
  differential operators. In particular, popular discretizations such
  as pseudospectral discretizations using the fast Fourier transform
  can be transformed to gauge invariant schemes. Also generic
  gauge invariant versions of generic time integration methods are
  considered, enabling completely gauge invariant calculations of the
  time dependent Schr\"odinger equation. Numerical examples
  illuminating the differences between a gauge invariant
  discretization and conventional discretization procedures are also
  presented.
\end{abstract}

\maketitle
%\newpage

\section{Introduction}
The fundamental laws of physics can (without exceptions) be related to
certain continuous symmetries. In other words, by requiring that a
model should be invariant with respect to a certain symmetry, the
model is more or less completely determined. The Standard model
of particle physics \cite{cite:peskin95, cite:Weinberg, cite:quigg97}
and Gravitation \cite{cite:wheeler73} are examples of such theories.

As an example, a model with a complex scalar field, i.e., a model of
charged bosons, and a requirement of local $U(1)$-invariance, or
gauge invariance, will immediately yield the Maxwell-Klein-Gordon
(MKG) theory, which in the non-relativistic limit reduces to
Maxwell-Schr\"odinger theory. In addition to defining the theory, the
continuous symmetries give rise to conserved quantities through
Noether's theorem(s) \cite{cite:olver00, cite:goldstein02,
  cite:rubakov02}, and the local $U(1)$-symmetry of the MKG-model
ensures the conservation of local electric charge.

In particle physics, and especially in the QCD-part of the standard
model, numerical calculations are often done using Lattice Gauge
Theory (LGT) \cite{cite:wilson74, cite:rothe05, cite:creutz86}. This
is a numerical procedure, actually motivated from the continuous
theory, designed to preserve the underlying continuous gauge
symmetry. In a previous article this discretization scheme was applied
to the MKG-equation, with emphasis on the continuous $U(1)$-symmetry
and conservation laws deduced from discrete versions of Noether's
theorem(s) \cite{cite:halvorsen08}. By preserving the $U(1)$-symmetry
of the MKG-model on the discrete level, a discrete equivalent of the
conservation of local electric charge is immediate, which not only
makes the scheme consistent, but is also a good indicator of
stability. By a more standard discretization of the model, the local
$U(1)$-symmetry is broken, which again implies that the scheme is not
consistent with the continuous formulation. This will also reveal
itself through the fact that the physical observables calculated are
dependent on the gauge chosen, obviously in conflict with the
continuous model.

A similar breaking of the continuous local $U(1)$-symmetry has been a
known issue with discretizations of the Schr\"odinger equation coupled
to an external electromagnetic field. For example in atomic physics,
results are known to depend on the gauge in which the calculation is
done \cite{Madsen2002} -- a most unfortunate situation indicating that
the calculations are not correct. Gauge dependence also leads to
interpretation problems of the results. It is the goal of the present
paper to show how gauge invariant discretizations may be built from
existing ones with little or no extra effort in the implementations.
 
Simple gauge invariant grid discretizations of Schr\"odinger operators
have been studied in previous articles in the LGT formalism with
promising results \cite{cite:governale,cite:vogl08,cite:janecek}. The
key to success in LGT is that it does not approximate the covariant
derivative as a linear combination of the gradient and the gauge
potential, an element of the Lie-algebra under consideration, since
such an approximation leads to non-local terms when discretizing the
gradient, and a question of gauge invariance is meaningless since one
compare fields at different spacetime points. Instead LGT uses
Forward-Euler/central difference approximation of the gradient in the
various directions, which as argued is not gauge invariant, and then
defines the covariant derivative through the way non-local terms are
made gauge invariant in the continuous theory. This is done via the
Wilson line \cite{cite:wilson74, cite:peskin95}, to be discussed in
the next section, which effectively localize non-local terms by
parallel transport with the gauge potential as a key ingredient. By
defining the covariant derivative in this way, the discrete theory is
immediately manifestly gauge invariant.

The aim of this article is to expand the LGT formulation to allow for
completely general grid discretizations in arbitrary local coordinates
of the spatial manifold. Grid discretizations are widely used, and
include most numerical discretizations of Schr\"odinger operators in
use today, such as pseudospectral methods based on the discrete
Fourier transform or Chebyshev polynomials. We also generalize the
discussion to arbitrary coordinate systems, and some care is needed in
case some of the coordinates are periodic when using global
approximations (e.g., Chebyshev or Fourier expansions).

% The paper is organized as follows: In Section \ref{sec:schroedinger}
% we introduce the time dependent Schr\"odinger equation in
% general coordinates. In Section \ref{sec:discretization} we discuss
% gauge invariant (spatial) grid discretizations, while we in Section
% \ref{sec:time} discuss time evolution of the semidiscrete
% equations. Finally, in Section \ref{sec:results} we compare some
% gauge invariant scheme versions of standard schemes, before we close with some
% concluding remarks in Section \ref{sec:conclusion}.
% \\
% %--------------------------------------------
% \\
% Alternatively \\

The paper is organized as follows: In Section \ref{sec:schroedinger}
we introduce the time dependent Schr\"odinger equation in
general coordinates. In Section \ref{sec:discretization} and \ref{sec:covariant} we discuss
gauge invariant spatial grid discretizations. We proceed in Section
\ref{sec:time} to discuss gauge invariant time integration. Finally, in Section \ref{sec:results} we present some numerical results shedding light on the difference between gauge invariant and gauge dependent schemes, before we close with some concluding remarks in Section \ref{sec:conclusion}.

\section{The time dependent Schr\"odinger equation and gauge
  invariance}
\label{sec:schroedinger}

We consider a particle with charge $q$ and mass $m$ coupled to an
external electromagnetic (EM) field \cite{cite:shankar}
$(\mathbf{E},\mathbf{B})$. This is a semiclassical approach because
the EM-field is obviously affected by the particle, but if we assume
that the coupling is weak the approximation can be justified. We will
work in the non-relativistic regime, but our considerations could
easily be transmitted to a relativistic model. Moreover, the
generalization to more than one particle is straightforward, since the
EM fields only enter a many-body Hamiltonian at the one-body level,
i.e., the interparticle interactions are independent of the EM fields.

We are considering a spacetime domain $\mathbb{R}\times \Man$, with
coordinates $(t,y)$, where $t\in\mathbb{R}$ is the time coordinate,
and $y\in\Man$ is a point in the spatial domain, usually taken to be
Euclidean space, but can in general be a Riemannian manifold. In any
case, we may work in local coordinates $x=(x^i)$, viz,
$y=y(x)\in\Man$, with the induced metric tensor $g_{ij}(x)$ assumed to be time-independent. 
The wavefunction at some time $t$ is then a complex valued scalar function
$x\mapsto \psi(x)$. In addition, the EM-field is described by a gauge
potential $(t,x)\mapsto \phi(t,x)dt + A(t,x)$, where $\phi$ is a real
valued function and $A$ is a real valued one-form. In coordinate basis 
one usually identifies one-forms with vectors. Thus, if $\{dx^i\}$ are 
basis one-forms and $\{\mathbf e_i\}$ are basis vectors there is a 
one-to-one correspondence between $A = A_idx^i$ and 
$\mathbf A = A^i\mathbf e_i$. Note, we use the Einstein summation 
convention except where noted. The components of $A$ and $\mathbf A$
are related by the metric, i.e. $A^i = g^{ij}A_j$, and the physical EM fields
$(\mathbf{E},\mathbf{B})$ are given by
\begin{equation}\label{cont:EM_field}
\mathbf E = -\nabla \phi - \partial_t \mathbf A, \qquad \mathbf B = \curl \mathbf A,
\end{equation} 
where we use the shorthand $\partial_t = \partial/\partial t$.

In the following we work in units such that $\hbar = 1$. The dynamics of 
the system is governed by the time dependent
Schr\"odinger equation reading
\begin{equation}
  \label{eq:schroedinger}
  \I D_t\psi(t,x) = \left[-\frac{1}{2m}\Delta_{A} + V(t,x)\right]\psi(t,x) ,
\end{equation}
where $D_t = \partial_t + \I q\phi$ is the covariant derivative in the
temporal direction, and where the ``covariant Laplace-Beltrami''
operator $\Delta_A$ is defined by
\begin{equation}
  \Delta_A \equiv \frac{1}{\sqrt{g(x)}}D_j\sqrt{g(x)}g^{jk}D_k,
  \label{eq:laplace-beltrami-A}
\end{equation}
with $D_k = \partial_k - \I qA_k$ being the covariant derivative in
the direction $k$.  Moreover, $p_k = -\I D_k$ is the
(generalized) canonical momentum operator. The term
$-\Delta_A/2m$ is simply the kinetic energy operator, and
$V(t,x)$ is an external potential.

A fundamental property of the time dependent Schr\"odinger equation is
that it is invariant under local gauge transformations, i.e. equation
\eqref{eq:schroedinger} is invariant under the following set of
transformations
\begin{eqnarray}
  \psi(t,x)&\mapsto& e^{\I q\lambda(t,x)}\psi(t,x), \\ 
  \phi(t,x)&\mapsto& \phi(t,x) -
  \partial_t\lambda(t,x), 
  \\ A(t,x)&\mapsto& A(t,x) + \nabla\lambda(t,x)\label{cont:gauge_tr}, 
\end{eqnarray}
where $(t,x)\mapsto \lambda(t,x)$ is a real valued function meaning
that $\I q\lambda(t,x)\in\mathfrak{u(1)}$, the Lie-algebra of $U(1)$
(consult e.g. \cite{cite:olver00, cite:helgason78} for theory on
Lie-groups and Lie-algebras). One says that the theory is invariant
under local $U(1)$-transformations, meaning that the physical
observables are not affected by the transformations. In particular, we
note that the electric and magnetic fields \eqref{cont:EM_field} are
not affected by the transformations \eqref{cont:gauge_tr}. Moreover,
if $X_A=X_A[(p^j),(x^j)]$ is an observable, then the expectation value
$\langle \psi, X_A\psi\rangle$ is gauge invariant, viz,
\begin{equation}
  \langle e^{\I q\lambda} \psi, X_{A+\nabla\lambda}
  e^{\I q\lambda}\psi\rangle = \langle \psi, X_A\psi\rangle,
\end{equation}
where $\langle\cdot,\cdot\rangle$ denotes the standard inner product
in $L^2(\Man)$.

The usual way to write Eqn.~(\ref{eq:schroedinger}) is
\begin{eqnarray}
  \label{eq:schroedinger2}
  \I  \partial_t\psi(t,x) &=& H(t)\psi(t,x) \\ &:=&
  \left[-\frac{1}{2m}\Delta_A + 
    V(t,x) +  q \phi(t,x)\right]\psi(t,x),\notag
\end{eqnarray}
where the Hamiltonian $H(t)$ depends on the fields $(A,\phi)$. It is
well-known that for any two $t$,$t'$, the formal solution to
\eqref{eq:schroedinger2} is given by $\psi(t,x) =
\mathcal{U}(t,t')\psi(t,x')$, where the propagator $\mathcal{U}$ is
\begin{equation}
  \mathcal{U}(t,t') = \mathcal{T}\exp\left(-\I\int_{t'}^t H(s)\, ds
  \right),
  \label{eq:propagator}
\end{equation}
with $\mathcal{T}$ being the standard time-ordering operator. The
propagator depends on the fields $(A,\phi)$ in the case of the current
Hamiltonian, and under a gauge transformation with parameter
$\lambda(t,x)$ we have
\begin{equation}
  \mathcal{U}_{A',\phi'}(t,t') = e^{\I q \lambda(t)}
  \mathcal{U}_{A,\phi}(t,t') e^{-\I q \lambda(t')},
  \label{eq:propagator-trafo}
\end{equation}
where $A' = A + \nabla\lambda$ and $\phi' = \phi - \partial_t\lambda$.

\section{Discretization on a spatial grid}
\label{sec:discretization}

\begin{figure}
%\begin{center}
\includegraphics{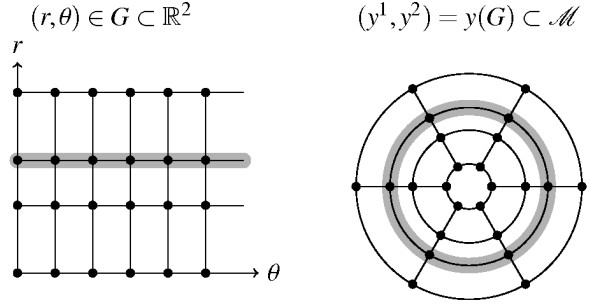}
%\end{center}
\caption{A grid in polar coordinates\label{fig:grid}. In local
  coordinates $(r,\theta)\in\mathbb{R}^2$ the grid $G$ is a Cartesian
  product, while in on the manifold $\Man\subset\mathbb{R}^2$ the grid instead
  has a (discrete) rotational symmetry. A coordinate curve for
  constant $r$ is also illustrated. In this case, the curve becomes a circle.}
\end{figure}

Many discretizations of Eqn.~\eqref{eq:schroedinger} approximate the
wave function $\psi(t,x)$ at a given time $t\in\mathbb{R}$ on a finite
grid $G$ (i.e., $y(G)\subset \Man$) in order to obtain a semi-discrete
formulation in which $\psi(t,\cdot)\in L^2[y(G)]$ still depends
continuously on time. We shall here consider grids which in the local
coordinates are Cartesian products of one-dimensional grids, i.e.,
\begin{equation}
  G = G^1 \times G^2 \times \cdots \times G^n,
  \label{eq:grid}
\end{equation}
where
\begin{equation}
  G^j = \{ x^j_1, x^j_2 , \cdots ,x^j_{N_j} \}, \quad x^j_k < x^j_{k+1}.
  \label{eq:grid2}
\end{equation}
Figure \ref{fig:grid} illustrates this in the case of polar
coordinates in the plane. 

We may list the elements of $G$ using
multi-indices, i.e.,
\begin{equation}
  G = \left\{ x_\alpha = (x^1_{\alpha_1},\cdots,x^n_{\alpha_n}) \;:\;
  \forall j,\;0 \leq \alpha_j < N_j \right\},
\label{eq:grid-list}
\end{equation}
and the multi-indices may again be mapped one-to-one with
$\{0,1,\cdots,N-1\}$, where $N=N_1N_2\cdots N_n$ is the total number
of grid points. Thus, we obtain a discrete Hilbert space
$\mathcal{H}(G)\simeq L^2[y(G)]$ of dimension $N$.

The natural basis to use in the space $\mathcal{H}(G)$ is the set of
functions $e_\alpha$ such that
$e_\alpha(x_\beta)=\delta_{\alpha\beta}$. These functions are referred
to as the cardinal basis \cite{cite:boyd01} or the nodal basis. For
any $\psi\in\mathcal{H}(G)$, we now have
\begin{equation}
  \psi = \sum_\alpha \psi(x_\alpha)e_\alpha.
\end{equation}
A linear operator $X$ on
$\mathcal{H}$ may be represented by its action on this basis, which
determines an $N\times N$ matrix with elements $X_{\alpha\beta}$, viz,
\begin{equation}
  X_{\alpha\beta} \equiv [Xe_\alpha](x_\beta).
\end{equation}
Thus, for $\psi\in\mathcal{H}(G)$,
\begin{equation}
  [X\psi](x_\alpha) = \sum_\beta X_{\alpha\beta}\psi(x_\beta).
\end{equation}
At times, we will omit the brackets and write $X\psi(x_\alpha)$ for
the product $X\psi$ evaluated at $x_\alpha$, as usually there is no
danger of confusion. Likewise, multiplication by $u\in\mathcal{H}(G)$
(or any continuous $u$ over $\Man$) defines a linear operator, and for
$\psi\in\mathcal{H}(G)$ we will write this simply as $u\psi$.

The grid discretization invariably comes with a discrete
approximation to the (field-free, i.e., $A=0$)
Laplace-Beltrami operator $\Delta_0$, although the exact procedure to
fix this discrete operator may vary. We may assume that this
discretization is composed of discrete derivatives in each spatial
direction $x^j$ composed with some fixed functions of the coordinates,
but the exact form is of no consequence to us.

To clarify these statements, consider for example Cartesian coordinates
in two spatial dimensions for which $\Delta_0 = \partial_x^2
+ \partial_y^2$. In a finite difference approximation we typically have a standard
5-point central difference stencil, i.e.,
\begin{equation}
  \Delta_0\psi(x^1,x^2) \approx -(\delta_1^\dag\delta_1 +
  \delta_2^\dag\delta_2)\psi(x^1,x^2), 
\end{equation}
where $\delta_j$ is a forward difference, $-\delta_j^\dag$ a backward
difference, so that $-\delta_j^\dag\delta_j$ is the standard 3-point
central difference operator in the $x^j$-direction, viz,
\begin{equation}
  -\delta_j^\dag\delta_j f(x^j) \equiv \frac{1}{h^2}[f(x^j+h) -
  2f(x^j) + f(x^j-h)], 
  \label{eq:stencil}
\end{equation}
with $h$ being the mesh width. We have suppressed other spatial
coordinates than $x_j$ in the latter equation.

As a different example, consider polar coordinates $(x^1,x^2) =
(r,\theta)$ in two dimensions,
for which
\begin{equation}
  \Delta_0 = \frac{1}{r}\partial_r r \partial_r +
  \frac{1}{r^2}\partial_\theta^2.
\end{equation}
The form of the radial part of this operator leads initially to
several different schemes by either expressing it
as
\begin{equation}
  \frac{1}{r}\partial_r r \partial_r = \partial_r^2 + \frac{1}{r}\partial_r
\end{equation}
and \emph{then} discretizing, or instead attack the original
difference operator. In a general coordinate system there will of
course be even more possibilities.

In any case, the discrete Laplace-Beltrami operator will be on the
form
\begin{equation}
  \Delta_0 \approx \Delta_{0,h} =
  \Delta_{0,h}(\delta_{j},\delta_{j}^2,x),
\end{equation}
where $\delta^k_{j}$ are arbitrary approximations to each partial
derivative $\partial^k_j$. We abuse notation a little, as in general
we allow $\delta^k_j \neq (\delta^1_j)^k$. In the Cartesian coordinate
example above, $\delta^2_j = -\delta_j^\dag\delta_j$. In general,
however, we assume that like in this example, $\delta_j^k$ is the
product of $k$ discrete derivative operators $\delta_{j,\ell}$,
$1\leq\ell\leq k$, so that
\begin{equation}
  \Delta_0 \approx \Delta_{0,h} =
  \Delta_{0,h}(\delta_{j,k},x),
\end{equation}
with $\delta_{j,0} = \delta_j$.

The usual way to discretize $\Delta_A$, on the other hand, which we
here will call a ``na\"ive'' discretization, is to employ a similar
``recipe'' as in the examples to Eqn.~\eqref{eq:laplace-beltrami-A}
after insertion of $D_j = \partial_j - iqA_j$ and simplifying the
expression. This, however, \emph{always} leads to non-gauge invariant
discretizations, as we will discuss in Section \ref{sec:covariant}.

As an example of the na\"ive approach, consider again the polar
coordinate case, and for simplicity assume $A_r = 0$ for which we
obtain
\begin{eqnarray*}
  \Delta_A &=& \frac{1}{r}\partial_r r \partial_r + 
   \frac{1}{r^2}(\partial_\theta -\I qA_\theta)^2 \\
  &=& \Delta_0 - \I q\frac{1}{r^2} (\partial_\theta A_\theta +
  A_\theta\partial_\theta) - \frac{q^2}{r^2}A_\theta^2.
\end{eqnarray*}
Assuming further that $\partial_\theta A_\theta = 0$, i.e., that $A$
is given in the Coulomb gauge, we get
\begin{equation}
  \Delta_A = \Delta_0 - \I 2 q\frac{1}{r^2}A_\theta \partial_\theta -
  \frac{q^2}{r^2}A_\theta^2. \label{eq:naive-polar}
\end{equation}
The na\"ive discretization of Eqn.~\eqref{eq:naive-polar} is then
given by inserting the usual grid discretizations of $\partial_r$,
$\partial_r^2$ and $\partial_\theta^2$.

Our prescription for a manifestly gauge invariant discretization of
$\Delta_A$ in Eqn.~\eqref{eq:laplace-beltrami-A} is simply to replace
\emph{all} occurrences of approximations $\delta_{j,k}$ of $\partial_j$
in the field-free na\"ive discretization with a certain corresponding
approximation $\tilde{D}_{j,k}$ to $D_j$, derived using methods from
LGT as mentioned in the Introduction, and whose final expression is
given in Eqn.~(\ref{eq:global-discrete-cov2}) below. In other words,
\begin{equation}
  \Delta_A \approx \Delta_{0,h}(\tilde{D}_{j,k},x),
\end{equation}
which will be gauge invariant. This approximation is often quite
different from the standard na\"ive discretization.

\section{Definition of the discrete covariant derivative}
\label{sec:covariant}

\subsection{One dimensional manifolds}
\label{sec:one-dim}

\subsubsection{Gauge transformations}
\label{sec:one-dim-gauge}

Consider first the case when $\Man$ is a
one-dimensional manifold $\Man=\Man^1\subset\mathbb{R}^n$. The reason
for this is that in the general case, the differentiation operator
$\partial_j$ can be viewed as a differential operator on the
coordinate curves, these being one-dimensional manifolds. Similarly,
a generic discrete $\delta_j$ can be viewed as an operator acting on
grid functions over the one-dimensional ``coordinate grids'' obtained
by fixing all but the $j$'th component of the multi-index $\alpha$ in
Eqn.~\eqref{eq:grid-list}. Equivalently, $\delta_j$ defines a discrete
differentiation operator acting on functions over a discretization of
the coordinate curve; see Fig.~\ref{fig:grid}. 

Any one-dimensional manifold $\Man$ will \emph{either} be
topologically equivalent to a circle \emph{or} an interval, which may
be bounded or unbounded. For example, in polar coordinates
$(x^1,x^2)=(r,\theta)$ in $\mathbb{R}^2$ the angular coordinate curves are circles of
radius $r$ while the radial coordinate curves are rays from the origin $r=0$
to infinity with an angle $\theta$ relative to the $x$-axis.

We write $x=x^1$ for the sole coordinate, omit the time dependence, and $D_A = \partial_x - \I
qA(x)$ for the covariant derivative. Under gauge transformations,
$D_A$ transforms as
\begin{equation}
  D_{A+\lambda'} = \partial_x - \I q[A(x) + \lambda'(x)] =
  e^{\I q\lambda(x)} D_A e^{-\I q\lambda(x)},
\end{equation}
where $\lambda'(x)=\partial_x\lambda(x)$. Intuitively, since $\Man$ is
one-dimensional, one should be able to transform away $A(x)$
completely, by selecting $\lambda' = -A$. However, if $\Man$ is
(topologically) a circle (with the point $x=0$ identified with $x =
L$, for simplicity), this is not possible: There are one-forms $A(x)$
which are \emph{not} the derivative of some zero-form $\lambda(x)$. On
the circle, it is precisely the constant functions $A(x) = A_0$, since
then $\lambda(x) = A_0x + b$ is \emph{not} a zero-form: it is not
periodic in $x$ unless $A_0=0$! If, on the other hand, $\Man$ is topologically an
interval, $A(x)$ may be transformed away.

These considerations may become clearer when we observe that, locally,
we may write
\begin{equation}
  D_A = u(x)^*\partial_x u(x), 
\end{equation}
where
\begin{equation}
u(x) = \exp\left(-\I q\int^x
    A(s)\,ds \right).
\end{equation}
Whenever $\Man$ is topologically an interval, we can choose $\lambda' = -A$, and this expression is
global, since then
\begin{equation}
  u(x) = \exp(\I q \lambda(x)).
\end{equation}
For $\Man$ being topologically a circle, no such $\lambda(x)$ exists
globally, unless $A_0=0$.

\subsubsection{Local approximations}

Let $\mathcal{H}(G)$ be the discrete Hilbert space corresponding to an
$N$-point discretization of $\Man=\Man^1$, being either a circle or an
interval as described above. Thus, $G\subset \Man^1$ is given by
\begin{equation}
  G = \{ x_1, x_2, \cdots x_N \}, \quad x_k< x_{k+1}.
\end{equation}
In the case of a circle, we identify $x_{N+1}$ and $x_1$ to impose
periodic boundary conditions. We let $h=\min(x_k,x_{k+1})$ be the mesh
width, and typically $h\sim 1/N$.

Let $\delta_h$ be a discrete differential operator on $\mathcal{H}(N)$,
and we assume for the moment that $\delta_h$ is a \emph{local}
operator, in the sense that as $h\rightarrow 0$, only a finite number
of points in the neighborhood of $x_k \in G$ are used to
differentiate $\psi(x_k)$. As a consequence, there is a largest
$p>0$ such that for any smooth function $\psi(x)$ over $\Man^1$, 
\begin{equation}
  \delta_h\psi(x_k) = \partial_x\psi(x_k) + O(h^p),
  \label{eq:approx-order}
\end{equation}
where the term $O(h^p)$ is equal to the truncation error, and we say that
$\delta_h$ is a $p$'th order approximation. Examples of local
discretizations are finite differences of any order, but not
pseudospectral methods using for example Chebyshev polynomials or the
discrete Fourier transform.

For any (discrete or smooth) $\psi$, the na\"ive discretization
$\hat{D}_{A,h}$ of $D_A$ reads
\begin{equation}
  \hat{D}_{A,h}\psi(x_k) = (\delta_h - \I qA)\psi(x_k). 
\end{equation}
This operator is {not} gauge invariant. Let $\lambda(x)$ be given,
and consider 
\begin{eqnarray*}
  (\hat{D}_{A+\lambda',h}e^{i\lambda}\psi)(x_k) &=& (\delta_h e^{\I q \lambda}\psi)(x_k) - \I q (A e^{\I q \lambda}\psi)(x_k) \\ &\neq& e^{\I q \lambda}
  \hat{D}_{A,h} \psi(x_k).
\end{eqnarray*}
Clearly, this comes about since 
\begin{equation*}
  (\delta_he^{\I q\lambda})(x_k) = \I q \lambda'(x_k)\exp(\I q\lambda(x_k))
  + O(h^p)
\end{equation*}
is only an approximation. 

However, the continuous covariant derivative $D_A$ comes about if
one tries to construct a gauge invariant classical field theory
\cite{cite:peskin95}: since for any differentiable $\psi$, 
$\psi(x)$ and $\psi(y)$ transforms differently if
$x\neq y$, the
limit
\begin{equation}
  \lim_{h\rightarrow 0} \frac{1}{h}[\psi(x+h)-\psi(x)]
\end{equation}
has no simple transformation law. Notice that for finite $h$,
$[\psi(x+h)-\psi(x)]/h$ is the standard forward difference
operator. We could just as well consider the limit
\begin{equation}
  \lim_{h\rightarrow 0} \delta_h\psi(x)
\end{equation}
for any local discrete differentiation operator.

Introducing a comparator function $U(x,y)$ with the transformation law
\begin{equation}
  U(x,y) \longrightarrow e^{\I q\lambda(x)}U(x,y)e^{-\I q\lambda(y)},
\end{equation}
we see that for any $y$, the function $U(x,y)\psi(y)$ transforms in
the same way as $\psi(x)$. Explicitly, the comparator is given by
\begin{equation}
  U(x,y) = e^{-\I q \int^{y}_{x} A(t) \, dt}.
\end{equation}
For any finite $h$, consider again the discrete difference operator
$\delta_h$ applied to $U(x,y)\psi(y)$, but acting on the variable $y$, i.e.,
\begin{equation}
  \tilde{D}_h\psi(x_k) \equiv [\delta_{h,y} U(x_k,y)\psi(y) ](x_k).
  \label{eq:local-discrete-cov}
\end{equation}
The notation implies that the discrete derivative is evaluated at
$y=x_k$. This operator is obviously gauge invariant, so that
\begin{equation}
  \tilde{D}_{A+\lambda',h}(e^{i\lambda}\psi)(x_k) = e^{i\lambda}
  \tilde{D}_{A,h} \psi (x_k).
\end{equation}

The path from $x$ to $y$ is in general ambiguous if $\Man$ is a
circle: we may move either clockwise or anti-clockwise, and also
through several revolutions before ending up at $y$.  However, as
$h\rightarrow 0$, we desire our discrete $\tilde{D}_{A,h}$ to converge
to $D_A$. The truncation error does not vanish unless we choose the
\emph{shortest} path, with vanishing length.
Then, Eqn.~\eqref{eq:approx-order} implies that for any smooth $\psi(x)$,
\begin{equation}
  [\tilde{D}_{A,h}\psi](x) = D_A\psi(x) + O(h^p).
\end{equation}

\subsubsection{Global approximations}

The fact that $\delta_h$ was a \emph{local} approximation to the
derivative was crucial above, as it allowed us to resolve path
ambiguity. For a \emph{global} approximation this is
not the case.

A global approximation $d_h$ to $\partial_x$ in general has
\emph{exponential} order of approximation as it utilizes \emph{all}
grid points $x_k\in G$ to estimate the derivative. That is, for any smooth
$\psi(x)$, the truncation error is $O(h^{N}) = O(h^{1/h})$, i.e.
\begin{equation}
  d_h\psi(x) = \partial_x\psi(x) + O(h^{N}),
\end{equation}
so that the order of approximation in fact increases as $h\rightarrow
0$. 

As $U(x,y)$ may depend on path, and as $x$ and $y$ have arbitrary
separation in a global method, the limit $h\rightarrow 0$ does not
resolve the path ambiguity.

The only way to overcome this, is to ensure that the comparator itself
is path-independent. This is the case if and only if $A(x)
= \partial_x \Lambda(x)$ for some smooth $\Lambda$, since then the
fundamental theorem of analysis yields
\begin{equation}
  \int_{x'}^{x} A(t)\, dt = \Lambda(x) - \Lambda(x'),
\end{equation}
independently of the path taken.
For $\Man$ being a circle, this means
that $A(x)$ must be the derivative of a \emph{periodic} function.

We therefore decompose the one-form $A(x)$ as
\begin{equation}
  A(x) = A_0 + A_1(x),
  \label{eq:decomp1d}
\end{equation}
where $A_0$ is a constant such that $A_1(x)=\partial_x\Lambda(x)$ for
some smooth $\lambda(x)$. Formally, $A_0$ is the projection of $A(x)$
onto the orthogonal complement of the range $\partial_x$, i.e.,
$\Ran(\partial_x)^\bot$. The decomposition \eqref{eq:decomp1d} is unique
and it always exists. We may say that $A_1$ is the ``largest part of
$A$ that may be transformed away.'' We then obtain
\begin{equation}
  D_A = e^{\I q\int^x A_1} (\partial_x - \I q A_0)e^{-\I q\int^x A_1}
  \label{eq:with-constant}
\end{equation}
for the covariant derivative. It is clear that if $A_0$ is nonzero, it
may never be transformed away using a gauge transformation.

In the case of $\Man$ being an interval,
$\Ran(\partial_x)^\bot=\{0\}$ , implying $A_0 = 0$, and we get
\begin{equation}
  D_A = e^{\I q\int^x A(t) \, dt} \partial_x e^{-\I q\int^x A(t) \,
    dt},
\end{equation}
where any anti-derivative of $A(x)$ may be used. For $\Man$ being a
circle, however, 
\begin{equation}
 \Ran(\partial_x)^\bot=\{\text{ constant functions }\},
\end{equation}
since $\int A_0 = A_0 x + b$ is not periodic unless $A_0=0$. 
It is straightforward to show that
\begin{equation}
  A_0 = \langle A \rangle = \frac{1}{L} \int_0^L A(t)\, dt.
  \label{eq:A_0}
\end{equation}

We define a modified path independent comparator given by
\begin{equation}
  U(x,y) = \exp\left[-\I q \int^{y}_{x} (A(s)-A_0)\; ds \right].
\end{equation}
Since it is independent of path, we may write
\begin{equation}
  U(x,y) = u^*(x)u(y), \quad u(x)\equiv U(x_0,x),
  \label{eq:path-indep-comp}
\end{equation}
where $x_0 \in G$ is any reference point. Combining
Eqns.~(\ref{eq:with-constant}) and (\ref{eq:local-discrete-cov}) we
get
\begin{equation}
  \tilde{D}_{A,h}\psi(x_k) \equiv [\delta_{h,y} U(x_k,y)\psi(y)
  ](x_k) - \I q A_0 \psi(x_k),
  \label{eq:global-discrete-cov}
\end{equation}
and using Eqn.~\eqref{eq:path-indep-comp} we may rewrite this as
\begin{equation}
  \tilde{D}_{A,h}\psi(x_k) \equiv u^*\delta_{h} u \psi(x_k) - \I q A_0 \psi(x_k),
  \label{eq:global-discrete-cov2}
\end{equation}
which may be a more practical expression to implement. This covariant
derivative is valid for any one-dimensional manifold topology and any
discretization of the derivative, and we note that in particular for
$A_0=0$, it is equivalent to the original expression
\eqref{eq:local-discrete-cov} for local discrete derivatives.

\subsection{General manifolds}
\label{sec:general-manifolds}

For global methods, the one-dimensional case necessitated the
computation of $A_0$ given by Eqn.~\eqref{eq:A_0}. As the covariant
derivative in this case was gauge equivalent to using the na\"ive
discretization with a constant $A(x)=A_0$, one may wonder what we have
to gain from the approach in this case: Why not use the standard
na\"ive discretization using this particular, and physically
equivalent, gauge? Most manifolds are, however, not
one-dimensional. In this section, the case of $\Man=\Man^n$ being a
general $n$-dimensional manifold is treated by simply defining
$\tilde{D}_{A,h}$ for each spatial direction. In this case it is in
general \emph{not} true that the method becomes gauge equivalent to a
na\"ive discretization: we may not find an $A$ such that the problem
may be solved with a na\"ive discretization. Said in another way, on
$\Man^n$, the splitting \eqref{eq:decomp1d} becomes
\begin{equation}
  A(x) = A_0(x) + \nabla\Lambda(x),
  \label{eq:decomp}
\end{equation}
where $A_0 \in \Ran(\nabla)^\bot$ is the part of $A$ which may not be
transformed away, and this is of course not a constant function in
general.

In the $i$'th direction, at the point $y(x)\in\Man$, the continuous
covariant derivative is given by
\begin{equation}
  D(A)_i = \partial_i -\I q A_i(x),
\end{equation}
being an operator that constructs the $i$'th component of a one-form
field, i.e., $D(A)\psi(x)=[\nabla-\I qA(x)]\psi(x)$ is a one-form field with components
$D(A)_i\psi(x)$.

As discussed in Section \ref{sec:discretization}, we are given
discretized derivative operators $\delta_i$ (we suppress the subscript
``$h$'' in the sequel, and also the distinction between local and
global discrete differentiation operators $d_i$) which only involves
grid points along the $i$'th coordinate curve at $x_\alpha\in
G=G^1\times G^2\times\cdots\times G^n$; see Fig.~\ref{fig:grid} for an
illustration. Thus, $\delta_i$ may be viewed as a discrete derivative
on discretization of a one-dimensional manifold (the $i$'th coordinate
curve at $x_\alpha$) with grid $G^i$. From Section \ref{sec:one-dim},
we then have the discrete covariant derivatives $\tilde{D}(A)_i$ given
by
\begin{equation}
  \tilde{D}(A)_i \equiv u^*_i(x) \delta_i u_i(x) - \I q A_0(x)_i,
\end{equation}
where $A_0(x)_i$ is the quantity $A_0$ in Eqn.~(\ref{eq:decomp1d}) for
$D(A)_i$ -- in general not a constant since it depends on the other
coordinates $x^{j}$, $j\neq i$. Neither is it given by the
decomposition \eqref{eq:decomp}. Moreover, $u_i(x)$ (or more precisely
$u_i^*(x)u_i(y)$) is the corresponding path-independent comparator for
differentiation in the $i$'th direction.

To be precise, we write out these quantities in the general case.

The quantity $A_0(x_\alpha)_i$ is zero for coordinate curves that are
topological intervals, such as the radial coordinate curves in polar
coordinates. For periodic coordinates, such as the angle $\theta$ in
polar coordinates, the coordinate curves are circles. In that case, 
\begin{equation}
  A_0(x_\alpha)_i = \frac{1}{L_i(x_\alpha)}\int_{0}^{L_i}
  A_i(x^1,\cdots,x^{i-1},s,x^{i+1},\cdots) \; ds,
\end{equation}
where $L_i$ is the length of the coordinate curve. In polar
coordinates, $x_\alpha = (r_\alpha,\theta_\alpha)$, and $L_i = 2\pi
r_\alpha$.

The comparator function $u_i(x_\alpha)$ is now given by
\begin{equation}
  u_i(x_\alpha) = \exp\left(-\I q \int_{0}^{x_\alpha^i}
    \left(A_i(\cdots,s,\cdots) - A_0(x_\alpha)\right) \; ds  \right),
\end{equation}
where the arbitrary reference point has been chosen as $x_\alpha^i =
0$.

Gauge invariance of $\tilde{D}(A)_i$ follows from the gauge invariance
in the one-dimensional case. Clearly, gauge invariance necessitates
calculating the comparator functions and $A_0(x)_i$. However, these
enter the discretizations only as multiplicative operators which are
diagonal in the nodal basis. It is therefore a one-time calculation inducing little overhead in general.

\section{Time discretization}
\label{sec:time}

Thus far we have studied a discretization of the time-dependent
Schr\"odinger equation in continuous time. However, when solving the
problem numerically one needs to discretize the model in time as
well. Again, with inspiration from classical LGT this can be done
manifestly gauge invariant for every scheme with a grid based
approximation of the time derivative.

The standard way to propagate the Schr\"odinger equation
(\ref{eq:schroedinger}) is to attack the form \eqref{eq:schroedinger2}
instead, viz, 
\begin{equation}
  \I  \partial_t \psi(t,x) = \left[-\frac{1}{2m}\Delta_A + V(t,x) + q\phi(t,x)\right]\psi(t,x),
  \label{eq:schroedinger-noncovariant}
\end{equation}
and then use standard techniques to integrate, analogously to the
na\"ive spatial discretizations. However, this will of course lead to
non-gauge invariant solutions.

Let us consider how gauge invariant formulations of some simple
schemes can be constructed. For simplicity, we will assume that the
wave function $\psi(t,x)$ is only sampled at equally spaced points in
time, i.e., $t_n = n\tau$ with $n=0,1,\ldots$. At each time $t_n$, we
write $\psi^n\in\mathcal{H}(N)$ for the corresponding spatially
discrete wave function.

Let $\delta_t = \partial_t + O(\tau^k)$ be a local approximation, and
assume that a na\"ive discretization of a generic Schr\"odinger
equation with Hamiltonian $H(t)$ is given by
\begin{equation}
  \I\delta_t \psi^n = \sum_{|j|\leq m} c_j H^{n+j}\psi^{n+j}, 
  \label{eq:simple-scheme-naive}
\end{equation}
where $H^{n}$ is the Hamiltonian $H(t)$ evaluated at $t=n\tau$. Here,
$c_j$ are constants, the notation indicating that only a finite number
of such constants are involved. Such schemes include the standard
implicit Crank-Nicholson and leap-frog schemes \cite{Askar1978}. For
the Crank-Nicholson scheme, $\delta_t$ is the forward Euler
discretization, while $c_0 = c_1 = 1/2$ ($m=1$ and $c_{-1}=0$). For
the Leap-Frog scheme, $\delta_t$ is the centered difference with step
length $2\tau$ and $c_0 = 1$ (with $m=0$). Using similar
considerations as in Section \ref{sec:covariant} for spatial
differentiation operators, the corresponding gauge invariant
discretization of \eqref{eq:schroedinger} for arbitrary fields $\phi$
becomes
\begin{equation}
  \I\tilde{D}_t \psi^n = \sum_{|j|\leq m} \tilde{c}_{n,j} H_A^{n+j}\psi^{n+j}, 
  \label{eq:simple-scheme-gauge-invariant}
\end{equation}
where $\tilde{c}_{n,j} = c_j U(t_n,t_{n+j})$, with
\begin{equation}
  U(t,t')=\exp\left(\I\int^{t'}_t \phi(s)ds\right)
\end{equation}
being the comparator for the time coordinate. The Hamiltonian $H_A(t)$
is given by
\begin{equation}
  H_A(t) = -\frac{1}{2m}\Delta_A + V(t,x),
\end{equation}
and \emph{excludes} the term $q\phi(t,x)$ which is now absorbed
into the covariant derivative $\tilde{D}_t$.

We now observe something peculiar: The scalar field $\phi(t,x)$ may be
transformed away \emph{globally} by the gauge parameter $\lambda =
\int^t \phi(s)ds$, yielding the so-called temporal gauge. In this
gauge $U(t,t')=1$, so the gauge invariant scheme
\eqref{eq:simple-scheme-gauge-invariant} reduces to the na\"ive scheme
\eqref{eq:simple-scheme-naive} -- of course with a different
Hamiltonian $H_{A + \nabla\lambda}$. 

In fact, these considerations hold for \emph{any} gauge invariant
numerical integration scheme: gauge invariance implies that the
temporal gauge in particular may be used, for which the integration
method reduces to the na\"ive non-gauge invariant scheme applied to
$H_{A+\nabla\lambda}$. Notice, however, that the latter operator is
\emph{time dependent}, even if $V$ and the fields $A$ and $\phi$ are
time-independent functions.

To make this statement precise, let $\mathcal{U}_h(t+\tau,t)$ be a
general numerical propagation scheme for a generic Hamiltonian $H(t)$,
i.e., it approximates the propagator $\mathcal{U}(t+\tau,t)$ in
Eqn.~\eqref{eq:propagator}, viz,
\begin{equation}
  \mathcal{U}(t+\tau,t) = \mathcal{U}_h(t+\tau,t) + O(\tau^m),
\end{equation}
where $O(\tau^m)$ is the truncation error of the scheme. Thus, the wave
function $\psi^n$ is propagated by
\begin{equation}
  \psi^{n+1} = \mathcal{U}_h(t+\tau,t)\psi^n.
\end{equation}

Assuming that $\tilde{\mathcal{U}}_{A,\phi}$ is a gauge invariant
generalization of $\mathcal{U}_h$ applied to the Hamiltonian $H_A$, it
must transform according to Eqn.~\eqref{eq:propagator-trafo}.
The temporal gauge is achieved by selecting $\lambda = \Lambda$ given by
\begin{equation}
  \Lambda(t,x) = \int^t_0 \phi(x,s)\; ds,
\end{equation}
which gives
\begin{equation}
  A' = A + \nabla\Lambda(t,x) = A + \nabla \int^t_0 \phi(x,s)\; ds.
\end{equation}
We obtain
\begin{equation}
  \tilde{\mathcal{U}}_{A,\phi}(t,t') = e^{\I q \Lambda(t)}
  \tilde{\mathcal{U}}_{A+\nabla\Lambda,0}(t,t')e^{-\I q \Lambda(t')},
  \label{eq:prop-trafo2}
\end{equation}
where $\tilde{\mathcal{U}}_{A+\nabla\Lambda,0}$ must be equal to the
original gauge dependent propagator applied to the Hamiltonian
$H_{A+\nabla\Lambda}$.

It is not always easy to identify an expression for
$\tilde{\mathcal{U}}_{A,\phi}(t,t')$ in a general gauge, but from the
above considerations, the temporal gauge is sufficient anyway.

The selection of a particular gauge for time integration may seem
unnatural. However, it is the structure of the Schr\"odinger equation
together with the fact that any field $\phi$ may be transformed away
that yields this conclusion. The vector potential $A$ cannot in
general be transformed away -- therefore we should not choose a particular
gauge for spatial operators.

\section{Numerical example}
\label{sec:results}

In Ref.~\cite{cite:governale}, some promising gauge invariant
eigenvalue calculations are shown using the classical LGT formalism,
i.e., with standard finite differences in space. In
Ref.~\cite{cite:janecek} higher-order finite differences are used, and
the results are equally promising. Even though the published
experiments are all with the ``standard'' example of a uniform
magnetic field in the $z$-direction applied to a planar system, there
is little doubt that the gauge invariant formulations offer favorable
properties over the non-gauge invariant methods, as there are always
gauges that behave very badly. One may simply choose a rapidly
oscillating gauge parameter $\lambda(t,x)$ to completely destroy the
accuracy. Choosing the ``right gauge'' in a non-gauge invariant scheme
may not at all be simple or even possible. In any case, a
gauge independent method will ``factor out'' any non-physical effect
of the choice of gauge, making interpretations easier, and it is
reasonable to expect that gauge invariance should stabilize the
discretization because of this.

We focus here on time integration only. The benefit of employing spatially gauge invariant schemes has already been established in e.g. \cite{cite:governale}. A gauge invariant discretization of the time-dependent Schr\"odinger
equation could enable practitioners to push the limits of what is
possible to compute and interpret. 

We consider a single particle in a one-dimensional system; a very
simple system but one whose numerical properties are reflected in more
realistic settings. We set $m=q=1$, and consider the
Schr\"odinger equation \eqref{eq:schroedinger} on the form
\begin{equation}
  \I[\partial_t +\I \phi(t,x)]\psi(t,x) = -\frac{1}{2}[\partial_x -
  iA(t,x)]^2\psi(t,x).
  \label{eq:se1d}
\end{equation}
We consider a spatial truncation $[-L/2,+L/2]\subset\mathbb{R}$, and
use a finite difference discretization with $N+2$ equally spaced grid
points $x_k = k h \in G$, $k=0,1,\ldots,N+1$. The grid spacing is
given by $h=L/(N+1)$. Thus, at a time $t$, $\psi(t,x)\approx
\psi(t,x_j) \in\mathcal{H}(G)$ is our discrete wave function.

A common situation in atomic physics arise when one considers the
so-called dipole approximation \cite{Madsen2002}, in which the fields
$A$ and $\phi$ take the form
\begin{eqnarray}
  \phi(t,x) &=& f(t)x \notag \\
  A(t,x) &=& 0,
  \label{eq:length-gauge}
\end{eqnarray}
corresponding to a time-dependent electric field $\mathbf{E}(t) =
-f(t)\mathbf{e}_x$ and $\mathbf{B} = 0$. These fields are of course
not solutions of Maxwell's equations. The particular gauge in
Eqn.~\eqref{eq:length-gauge} is referred to as the length-gauge. The
so-called velocity gauge is obtained by transforming away $\phi(t,x)$
using the gauge parameter $\lambda(t,x) = \int^t_0 \phi(s,x) ds$,
i.e., it is the temporal gauge. We obtain
\begin{eqnarray}
  \phi'(t,x) &=& 0, \notag \\
  A'(t,x) &=& \int^t_0 \partial_x \phi(s,x)\, ds.
  \label{eq:velocity-gauge}
\end{eqnarray}
The wave functions in the two gauges are of course related by $\psi'(t,x) =
e^{i\lambda(t,x)}\psi(t,x)$. These two gauges are commonly studied,
and may give different results in actual calculations; a sure sign of
a significant error.

A common choice for $f(t)$ is
\begin{equation}
  f(t) = c \sin(\omega t),
\end{equation}
describing an oscillating electric field with frequency $\omega$.

Using finite differences, a typical na\"ive semi-discretization of
Eqn.~\eqref{eq:se1d} is
\begin{eqnarray}
  \I\partial_t\psi(t,x_j) &=& \left[-\frac{1}{2}(\delta^+ - \I A)(\delta^- -
  \I A) + \phi\right]\psi(t,x_j) \notag \\
&=& \frac{1}{2}\left[-\delta^+\delta^- + \I(A\delta^- +
  \delta^+ A) + A^2 + 2\phi\right]\psi(t,x_j).
  \label{eq:se1d2}
\end{eqnarray}
where $\delta^+$ is a forward difference, and $\delta^-$ is a backward
difference.  As earlier, $A$ and $\phi$ should be interpreted as
diagonal multiplication operators. As $(\delta^+)^\dag = -\delta^-$, it
is easy to see that the operator on the right hand side of
Eqn.~\eqref{eq:se1d2} is actually
Hermitian. 

The corresponding gauge invariant semi-discretization is, in the
temporal gauge,
\begin{equation}
  \I \partial_t\psi(t,x_j) = \left[-\frac{1}{2}u^*\delta^+\delta^-
    u\right]\psi(t,x_j),
  \label{eq:se1d3}
\end{equation}
with $u(t,x) = \exp[-\I \lambda(t,x) + \I\lambda(t,0)]$ being the
comparator function.

To integrate Eqns.~\eqref{eq:se1d2} and \eqref{eq:se1d3} in time, we
select a somewhat non-standard approach. It is well-known that an
approximation to $\mathcal{U}(t+\tau,t)$ for a given Hamiltonian
$H(t)$ is
\begin{equation}
  \mathcal{U}(t+\tau,t) \approx \mathcal{U}_\tau(t) \equiv e^{-i\int_t^{t+\tau} H(s) ds},
\end{equation}
where the error is $O(\tau^2)$. Propagating $\psi^n \equiv
\psi(t_n,x_j) \in \mathcal{H}(G)$ using
$\mathcal{U}_\tau(t)$ gives an error increasing roughly linearly as
function of the number of time steps. The integral is evaluated using
Gauss-Legendre quadrature using two evaluation points, giving
practically no error in the integral as long as $f(t)$ does not
oscillate too rapidly.

We choose $c=5$ and $\omega=10$ for the electrical field, and
integrate for $t_n = n\tau \leq 6\pi/\omega$, so that the electric
field oscillate exactly three times before we terminate the calculation. The spatial
domain is of length $L=40$, and we use $N=255$ points.

We choose $\psi(x,0) = \exp(-x^2/2)$ as initial condition (which is
normalized numerically in the calculations). The analytic solution
using this particular problem (on whole of $\mathbb{R}$) can be
computed in closed form, but we choose instead to perform a reference
calculation using a pseudospectral discretization using $N+1$ points
and a much smaller time step, giving in this case practically no
error.

\begin{figure}
%\begin{center}
\includegraphics{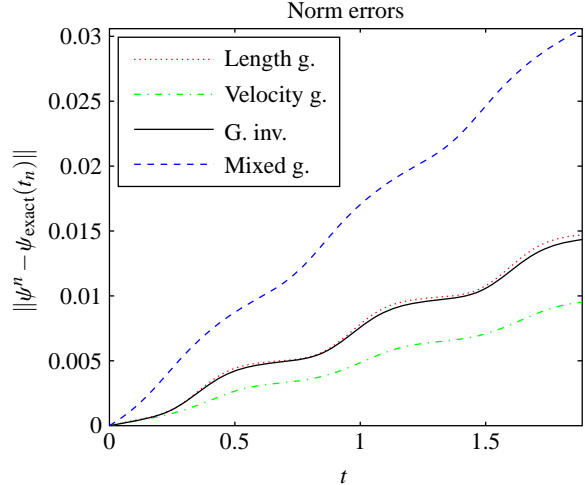}
%\end{center}
\caption{(Color online) Errors in the time-integration for the
  electric field $\mathbf{E}(t,x) = -c\sin(\omega
  t)\mathbf{e}_x$.\label{fig:numerical1} The velocity gauge
  (dashed/dotted green) has somewhat larger error than the 
  gauge invariant (solid black) and the length gauge (dotted red)
  calculations. The low error in the velocity gauge is a ``stroke of
  luck'' when compared with Fig.~\ref{fig:numerical2}, where the
  velocity gauge has the largest  error. Notice that the oscillations
  of the EM-fields clearly affect the errors.}
\end{figure}

Figure \ref{fig:numerical1} shows the error $\|\psi^n -
\psi_\text{exact}(t_n)\|$ as function of $t$ in the three
cases. Clearly, the velocity gauge has somewhat smaller error, and
also the length gauge and gauge invariant calculations have almost
indistinguishable errors. The latter fact can easily be understood by inserting
$\psi' = \exp(i\lambda)\psi$ into the semi-discrete formulations, and
noting that $\delta^+\delta^-$ and $u^*\delta^+\delta^-u$ are
unitarily equivalent, i.e., having the same eigenvalues. The
semidiscrete equations are thus actually equivalent, and any discrepancy
showing in the graphs for the gauge invariant scheme and the
length-gauge calculation comes from errors in the time integration.

The spectrum of $(\delta^+ -\I A)(\delta^- -\I A)$ is \emph{not}
equivalent to that of $\delta^+\delta^-$ when $A\neq 0$, however. In
fact, it is readily established that the latter operator has
eigenvalues depending strongly on $A$, and therefore on the particular
gauge used. Hence, it is expected that the velocity gauge, or any
other gauge in which $A\neq 0$, should perform \emph{worse} than
either of the other gauges in the generic case.

\begin{figure}
%\begin{center}
\includegraphics{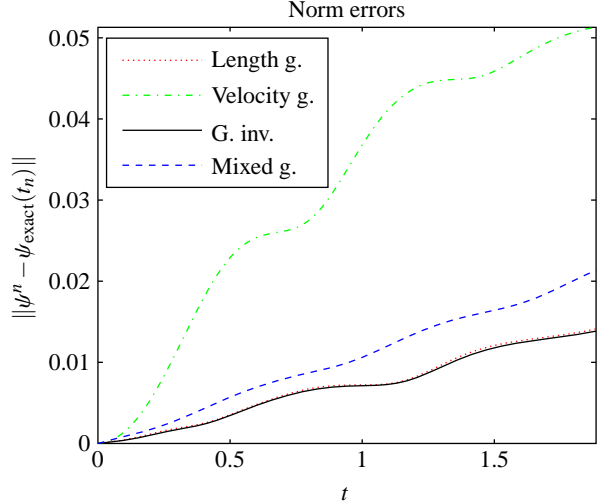}
%\end{center}
\caption{(Color online) Errors in the time-integration for the
  electric field $\mathbf{E}(t,x) = c\mu\cos(\omega t - \mu
  x)\mathbf{e}_x$.\label{fig:numerical2} The gauge invariant
  (solid black) and the length gauge (dotted red) calculations  have
  the smallest errors, while the velocity gauge (dot-dash, green) has
  clearly the largest error. This should be contrasted with the
  results in Fig.~\ref{fig:numerical1}, where the velocity gauge has the smallest error.}
\end{figure}

To test this statement, we perform a calculation using  a different
field $\phi(t,x)$ in the length gauge, namely
\begin{equation}
  \phi(t,x) = c \sin(\omega t - \mu x),
\end{equation}
which may describe incoming electromagnetic waves (e.g., a laser)
along the $x$ axis. Figure \ref{fig:numerical2} shows the errors as
function of $t$ in this case, using $\mu = 1$, clearly showing that the velocity
gauge indeed has the larger error. Moreover, a ``mixed gauge'' calculation
is shown, where the length gauge fields are transformed using a gauge
parameter $\lambda(t,x) = x$, chosen somewhat arbitrarily. Now, both
$A$ and $\phi$ are non-vanishing, and the error is seen to behave
accordingly.

We notice that the gauge invariant calculations in both cases are
well-behaved, and no choice of gauge will of course affect the
calculations. Moreover, the length gauge is equivalent to the
gauge invariant calculations \emph{only} when $A=0$; this holds in
general in one dimensional systems, but of course not in arbitrary
dimensions, where the magnetic field usually cannot be transformed
away in this way.

\section{Conclusion}
\label{sec:conclusion}

We have discussed a method based on LGT to convert virtually any
grid-based scheme for the time-dependent Schr\"odinger equation
(\ref{eq:schroedinger2}) to a gauge invariant scheme, in both space
and time. We have considered discretization in arbitrary coordinates
on arbitrary spatial manifolds. The theory is directly generalizable
to many-particle systems as the EM-fields obviously only enter at
one-body level in the many-body Hamiltonian. Moreover, the
computational overhead of the gauge invariant schemes compared to the
original ones are negligible.

Our numerical simulations of time-dependent problems, albeit
simplistic, indicate that the gauge invariant schemes perform on
average better than standard schemes, even though the original
``na\"ive'' scheme may be better in specific cases. 

A further line of work would be to rigorously understand the accuracy
gained by introducing gauge invariance.


\begin{thebibliography}{19}
\expandafter\ifx\csname natexlab\endcsname\relax\def\natexlab#1{#1}\fi
\expandafter\ifx\csname bibnamefont\endcsname\relax
  \def\bibnamefont#1{#1}\fi
\expandafter\ifx\csname bibfnamefont\endcsname\relax
  \def\bibfnamefont#1{#1}\fi
\expandafter\ifx\csname citenamefont\endcsname\relax
  \def\citenamefont#1{#1}\fi
\expandafter\ifx\csname url\endcsname\relax
  \def\url#1{\texttt{#1}}\fi
\expandafter\ifx\csname urlprefix\endcsname\relax\def\urlprefix{URL }\fi
\providecommand{\bibinfo}[2]{#2}
\providecommand{\eprint}[2][]{\url{#2}}

\bibitem[{\citenamefont{Peskin and Schroeder}(1995)}]{cite:peskin95}
\bibinfo{author}{\bibfnamefont{M.~E.} \bibnamefont{Peskin}} \bibnamefont{and}
  \bibinfo{author}{\bibfnamefont{D.~V.} \bibnamefont{Schroeder}},
  \emph{\bibinfo{title}{An introduction to Quantum Field Theory}}
  (\bibinfo{publisher}{Westview Press}, \bibinfo{year}{1995}),
  \bibinfo{edition}{1st} ed.

\bibitem[{\citenamefont{Weinberg}(2002)}]{cite:Weinberg}
\bibinfo{author}{\bibfnamefont{S.}~\bibnamefont{Weinberg}},
  \emph{\bibinfo{title}{The Quantum Theory of Fields}},
  vol.~\bibinfo{volume}{1} (\bibinfo{publisher}{Cambridge University Press},
  \bibinfo{year}{2002}).

\bibitem[{\citenamefont{Quigg}(1997)}]{cite:quigg97}
\bibinfo{author}{\bibfnamefont{C.}~\bibnamefont{Quigg}},
  \emph{\bibinfo{title}{Gauge Theories of the Strong, Weak, and Electromagnetic
  Interactions}} (\bibinfo{publisher}{Westview Press}, \bibinfo{year}{1997}).

\bibitem[{\citenamefont{{M}isner et~al.}(1973)\citenamefont{{M}isner, {K}ip
  {S}.~{T}horne, and {J}ohn~{A}rchibald {W}heeler}}]{cite:wheeler73}
\bibinfo{author}{\bibfnamefont{C.~W.} \bibnamefont{{M}isner}},
  \bibinfo{author}{\bibnamefont{{K}ip {S}.~{T}horne}}, \bibnamefont{and}
  \bibinfo{author}{\bibnamefont{{J}ohn~{A}rchibald {W}heeler}},
  \emph{\bibinfo{title}{Gravitation}} (\bibinfo{publisher}{W. H. Freeman and
  Company, New York}, \bibinfo{year}{1973}).

\bibitem[{\citenamefont{Olver}(2000)}]{cite:olver00}
\bibinfo{author}{\bibfnamefont{P.~J.} \bibnamefont{Olver}},
  \emph{\bibinfo{title}{Applications of Lie Groups to Differential Equations}}
  (\bibinfo{publisher}{Springer-Verlag}, \bibinfo{year}{2000}),
  \bibinfo{edition}{2nd} ed.

\bibitem[{\citenamefont{Goldstein et~al.}(2002)\citenamefont{Goldstein, Poole,
  and Safko}}]{cite:goldstein02}
\bibinfo{author}{\bibnamefont{Goldstein}},
  \bibinfo{author}{\bibnamefont{Poole}}, \bibnamefont{and}
  \bibinfo{author}{\bibnamefont{Safko}}, \emph{\bibinfo{title}{Classical
  Mechanics}} (\bibinfo{publisher}{Addison Wesley}, \bibinfo{year}{2002}),
  \bibinfo{edition}{3rd} ed.

\bibitem[{\citenamefont{Rubakov}(2002)}]{cite:rubakov02}
\bibinfo{author}{\bibfnamefont{V.}~\bibnamefont{Rubakov}},
  \emph{\bibinfo{title}{Classical Theory of Gauge Fields}}
  (\bibinfo{publisher}{Princeton University Press}, \bibinfo{year}{2002}),
  \bibinfo{edition}{1st} ed.

\bibitem[{\citenamefont{Wilson}(1974)}]{cite:wilson74}
\bibinfo{author}{\bibfnamefont{K.~G.} \bibnamefont{Wilson}},
  \bibinfo{journal}{Phys. Rev. D} \textbf{\bibinfo{volume}{10}},
  \bibinfo{pages}{2445} (\bibinfo{year}{1974}).

\bibitem[{\citenamefont{Rothe}(2005)}]{cite:rothe05}
\bibinfo{author}{\bibfnamefont{H.~J.} \bibnamefont{Rothe}},
  \emph{\bibinfo{title}{Lattice Gauge Theories, An Introduction}}
  (\bibinfo{publisher}{World Scientific}, \bibinfo{year}{2005}),
  \bibinfo{edition}{3rd} ed.

\bibitem[{\citenamefont{Creutz}(1986)}]{cite:creutz86}
\bibinfo{author}{\bibfnamefont{M.}~\bibnamefont{Creutz}},
  \emph{\bibinfo{title}{Quarks, gluons and lattices}}
  (\bibinfo{publisher}{Cambridge}, \bibinfo{year}{1986}),
  \bibinfo{edition}{1st} ed.

\bibitem[{\citenamefont{Christiansen and Halvorsen}(2008)}]{cite:halvorsen08}
\bibinfo{author}{\bibfnamefont{S.~H.} \bibnamefont{Christiansen}}
  \bibnamefont{and} \bibinfo{author}{\bibfnamefont{T.~G.}
  \bibnamefont{Halvorsen}}, \bibinfo{type}{Tech. Rep.},
  \bibinfo{institution}{University of Oslo, Department of Mathematics},
  \bibinfo{address}{E-print No. 17,Pure Mathematics, ISSN 0806-2439}
  (\bibinfo{year}{2008}).

\bibitem[{\citenamefont{Madsen}(2002)}]{Madsen2002}
\bibinfo{author}{\bibfnamefont{L.~B.} \bibnamefont{Madsen}},
  \bibinfo{journal}{Phys. Rev. A} \textbf{\bibinfo{volume}{65}},
  \bibinfo{pages}{053417} (\bibinfo{year}{2002}).

\bibitem[{\citenamefont{Governale and Ungarelli}(1998)}]{cite:governale}
\bibinfo{author}{\bibfnamefont{M.}~\bibnamefont{Governale}} \bibnamefont{and}
  \bibinfo{author}{\bibfnamefont{C.}~\bibnamefont{Ungarelli}},
  \bibinfo{journal}{Phys. Rev. B} \textbf{\bibinfo{volume}{58}},
  \bibinfo{pages}{7816} (\bibinfo{year}{1998}).

\bibitem[{\citenamefont{Andlauer et~al.}({2008})\citenamefont{Andlauer,
  Morschl, and Vogl}}]{cite:vogl08}
\bibinfo{author}{\bibfnamefont{T.}~\bibnamefont{Andlauer}},
  \bibinfo{author}{\bibfnamefont{R.}~\bibnamefont{Morschl}}, \bibnamefont{and}
  \bibinfo{author}{\bibfnamefont{P.}~\bibnamefont{Vogl}},
  \bibinfo{journal}{{Physical Review B}} \textbf{\bibinfo{volume}{{78}}}
  (\bibinfo{year}{{2008}}), ISSN \bibinfo{issn}{{1098-0121}}.

\bibitem[{\citenamefont{Janecek and Krotscheck}({2008})}]{cite:janecek}
\bibinfo{author}{\bibfnamefont{S.}~\bibnamefont{Janecek}} \bibnamefont{and}
  \bibinfo{author}{\bibfnamefont{E.}~\bibnamefont{Krotscheck}},
  \bibinfo{journal}{{Physical Review B}} \textbf{\bibinfo{volume}{{77}}}
  (\bibinfo{year}{{2008}}), ISSN \bibinfo{issn}{{1098-0121}}.

\bibitem[{\citenamefont{Shankar}(1994)}]{cite:shankar}
\bibinfo{author}{\bibfnamefont{R.}~\bibnamefont{Shankar}},
  \emph{\bibinfo{title}{Principles of Quantum Mechanics}}
  (\bibinfo{publisher}{Springer}, \bibinfo{year}{1994}), ISBN
  \bibinfo{isbn}{0306447908}.

\bibitem[{\citenamefont{Helgason}(1978)}]{cite:helgason78}
\bibinfo{author}{\bibfnamefont{S.}~\bibnamefont{Helgason}},
  \emph{\bibinfo{title}{Differential Geometry, Lie Groups and Symmetric
  Spaces}} (\bibinfo{publisher}{Academic Press}, \bibinfo{year}{1978}).

\bibitem[{\citenamefont{Boyd}(2001)}]{cite:boyd01}
\bibinfo{author}{\bibfnamefont{J.}~\bibnamefont{Boyd}},
  \emph{\bibinfo{title}{{Chebyshev and Fourier Spectral Methods}}}
  (\bibinfo{publisher}{Springer}, \bibinfo{year}{2001}).

\bibitem[{\citenamefont{Askar and Cakmak}(1978)}]{Askar1978}
\bibinfo{author}{\bibfnamefont{A.}~\bibnamefont{Askar}} \bibnamefont{and}
  \bibinfo{author}{\bibfnamefont{A.~S.} \bibnamefont{Cakmak}},
  \bibinfo{journal}{The Journal of Chemical Physics}
  \textbf{\bibinfo{volume}{68}}, \bibinfo{pages}{2794} (\bibinfo{year}{1978}).

\end{thebibliography}
\end{document}